\documentclass[letterpage, fleqn, usenatbib, useAMS]{mnras}
\usepackage{graphicx}
\usepackage{amsmath}
\usepackage{astrobib_mnras}
\usepackage{amssymb}
\usepackage{multicol}
\usepackage{multirow}
\usepackage{bm}
\usepackage{blkarray}
\usepackage{lineno}
\usepackage[T1]{fontenc}
\usepackage{ae,aecompl}
\usepackage{newtxtext,newtxmath}

\title[eBOSS DR14 quasar BAO with redshift weights]{The clustering of the SDSS-IV extended Baryon Oscillation
Spectroscopic Survey DR14 quasar sample: Measuring the anisotropic Baryon Acoustic Oscillations with redshift weights}

\author[Zhu et al]{\parbox{\textwidth}{Fangzhou Zhu$^1$,
Nikhil Padmanabhan$^1$,
Ashley J. Ross$^2$,
Martin White$^3$,
Will J. Percival$^4$,
Rossana Ruggeri$^4$, 
Gong-bo Zhao$^{4,5}$,
Dandan Wang$^{5}$,
Eva-Maria Mueller$^{4}$, 
Etienne Burtin$^{7}$,
H\'ector Gil-Mar\'in$^{8,9}$, 
Florian Beutler$^{4,6}$, 
Jonathan Brinkmann$^{10}$,
Joel R. Brownstein$^{11}$,
Kyle Dawson$^{11}$, 
Axel de la Macorra$^{12}$,
Graziano Rossi$^{13}$,
Donald P. Schneider$^{14, 15}$,
Rita Tojeiro$^{16}$, 
Yuting Wang$^{4,5}$
} \vspace*{4pt} \\
\scriptsize $^1$ Dept. of Physics, Yale University, New Haven, CT 06511, USA \\
\scriptsize $^2$ Center for Cosmology and Astroparticle Physics, Department of Physics, The Ohio State University, OH 43210, USA \\
\scriptsize $^{3}$ Dept. of Physics and Astronomy, U.C. Berkeley, Berkeley, CA 94720, USA \\
\scriptsize $^4$ Institute of Cosmology \& Gravitation, University of Portsmouth, Dennis Sciama Building, Portsmouth, PO1 3FX, UK \\
\scriptsize $^{5}$ National Astronomy Observatories, Chinese Academy of Science, Beijing, 100012, P.R.China \\
\scriptsize $^6$ Lawrence Berkeley National Lab, 1 Cyclotron Rd, Berkeley CA 94720, USA \\ 
\scriptsize $^7$ IRFU, CFA, Universit\'e Paris-Scalay, F-91191 Gif-sur-Yvette, France \\
\scriptsize $^8$ Sorbonne Universit\'es, Institut Lagrange de Paris (ILP), 98 bis Boulevard Arago, 75014 Paris, France \\
\scriptsize $^9$ Laboratoire de Physique Nucl\'eaire et de Hautes Energies, Universit\'e Pierre et Marie Curie, 4 Place Jussieu, 75005 Paris, France \\
\scriptsize $^{10}$ Apache Point Observatory, P.O. Box 59, Sunspot, NM 88349 \\
\scriptsize $^{11}$ Department of Physics and Astronomy, University of Utah, 115 S. 1400 E., Salt Lake City, UT 84112, USA  \\
\scriptsize $^{12}$ Instituto de F\'isica, Universidad Nacional Aut\'onoma de M\'exico, Apdo.Postal 20-364, 01000, M\'exico D.F., M\'exico \\
\scriptsize $^{13}$ Department of Physics and Astronomy, Sejong University, Seoul 143-747, Korea \\
\scriptsize $^{14}$ Department of Astronomy and Astrophysics, The Pennsylvania State University, University Park, PA 16802, USA \\ 
\scriptsize $^{15}$ Institute for Gravitation and the Cosmos, The Pennsylvania State University,
   University Park, PA 16802, USA \\
\scriptsize $^{16}$ School of Physics and Astronomy, University of St Andrews, North Haugh, St Andrews KY16 9SS, UK 
}

\begin{document}
\label{firstpage}
\pagerange{\pageref{firstpage}--\pageref{lastpage}}
\maketitle

\begin{abstract}

We present an anisotropic analysis of Baryon Acoustic Oscillation (BAO) signal from the SDSS-IV extended Baryon Oscillation Spectroscopic Survey (eBOSS) Data Release 14 (DR14) quasar sample. The sample consists of 147,000 quasars distributed over a redshift range of $0.8 < z < 2.2$. 
We apply the redshift weights technique to the clustering of quasars in this sample and achieve a 4.6 per cent measurement of the angular distance measurement $D_M$ at $z = 2.2$ and Hubble parameter $H$ at $z=0.8$.  
We parameterize the distance-redshift relation, relative to a fiducial model, as a quadratic expansion. The coefficients of this expansion are used to reconstruct the distance-redshift relation and obtain distance and Hubble parameter measurements at all redshifts within the redshift range of the sample. 
Reporting the result at two characteristic redshifts, we determine $D_M(z=1) = 3405\pm305 \  (r_{\rm d} / r_{\rm d, fid}) \ {\rm Mpc}$, $H(z=1) = 120.7\pm 7.3 \ (r_{\rm d,fid} / r_{\rm d}) \  {\rm km} \ {\rm s}^{-1}{\rm Mpc}^{-1}$ and 
$D_M(z=2) = 5325\pm249 \  (r_{\rm d} / r_{\rm d, fid}) \ {\rm Mpc}$, $H(z=2) = 189.9\pm 32.9 \ (r_{\rm d,fid} / r_{\rm d}) \ {\rm km} \ {\rm s}^{-1}{\rm Mpc}^{-1}$. These measurements are highly correlated. 
We assess the outlook of BAO analysis from the final quasar sample by testing the method on a set of mocks that mimic the noise level in the final sample. We demonstrate on these mocks that redshift weighting shrinks the measurement error by over 25 per cent on average. We conclude redshift weighting can bring us closer to the cosmological goal of the final quasar sample.
\end{abstract}


\section{Introduction}

Baryon acoustic oscillations (BAO) in the distribution of the galaxies are a powerful tool to map the expansion history of the universe via a `standard ruler' in galaxy clustering \citep{Sunyaev&Zeldovich1970, Peebles&Yu1970, Bond&Efstathiou1987, Hu&Sugiyama1996, Eisenstein&Hu}.
Pressure waves prior to recombination imprint a characteristic scale in the matter clustering at the radius of the sound horizon $r_d$ when the photons and baryons decouple shortly after recombination.  
The BAO manifests itself today in the two-point matter correlation function as an `acoustic peak' of roughly 150 Mpc. This feature of known length can be used as a standard ruler to constrain the distance-redshift relation and the expansion history of the universe. 

Different tracers of the underlying dark matter distribution have been used to successfully measure the peak. 
These analyses include galaxies \citep{DR12}, the Ly$\alpha$ forest \citep{Delubac2014, Bautista2017}, voids \citep{Kitaura2015}, and quasar-Ly$\alpha$ forest cross correlations \citep{Font-Ribera2014}. 
Since the first detection of BAO \citep{Cole2005, Eisenstein05} in the galaxy distribution over a decade ago, 
galaxy redshift surveys \citep{Blake2007, Kazin2010, Percival2010, Beutler2011, Padmanabhan2012, AndersonDR11, DR12} have been driving the measurement to ever increasing precision. Large surveys like Baryon Oscillation Spectroscopic Survey (BOSS) \citep{Dawson13, Alam15}, 
a part of the Sloan Digital Sky Survey III (SDSS-III) \citep{EisensteinDesign} 
have enjoyed great success in making per cent level cosmological distance measurements. 

The extended Baryon Oscillation Spectroscopic Survey (eBOSS) \citep{Dawson2016} is a new redshift survey within SDSS-IV \citep{Blanton2017}, the observations for which started in July 2014. 
The photometry was obtained on the 2.5-meter
Sloan Telescope \citep{Gunn2006} at the Apache Point Observatory in New Mexico, USA. As part of this program, eBOSS observes quasars that are selected to enable clustering studies. The quasar sample covers a redshift range of $0.8 < z < 2.2$. 
The final sample is forecasted to produce a 1.6 per cent spherically-averaged distance measurement \citep{Zhao2016}. This paper uses the DR14 quasar sample whose targeting and observation details are described in \citet{Abolfathi2018}. 

Samples from current and future generations of BAO surveys such as the eBOSS cover
a wide range of redshift. To improve the resolution of distance-redshift relation measurement, traditional BAO analyses
usually split the samples into multiple redshift bins and analyze the signals in these slices. 
One drawback of splitting the sample into multiple redshift bins is that the signal-to-noise ratio in each bin becomes lower, making the analysis more sensitive to the tails of the likelihood distribution. Furthermore, the signal across boundaries of disjoint bins is lost in such an analysis. 
While some of these disadvantages may be overcome by properly accounting for all the covariances among the slices, 
they add to the complexity of the analysis. 
There is also no consensus on how to optimally split the sample. 

To tackle the problems with binning outlined above, \citet{Zhu15} proposed 
using a set of redshift weights to compress the information in the redshift direction onto a 
small number of `weighted correlation functions'. Applying the redshift weights to the galaxy pair counts efficiently preserves nearly all the BAO information in the sample, leading to improved constraints of the distance-redshift relation parameterized in a simple generic form over the entire redshift extent of the survey. 
\citet{Zhu16} validated the redshift weighting method on BOSS DR12 galaxy mock catalogues that the weights afford tighter distance and Hubble measurements across the redshift range of a sample compared with the unweighted single-bin analysis. The method has also been demonstrated to produce robust and unbiased BAO measurements. 

This paper applies redshift weighting to the BAO analysis of the eBOSS DR14 quasar sample. These measurements complement the analysis in \citet{DR14alphabetical} and provide a first measurement of $H(z)$ from this sample. 
The paper is structured as follows: \S~\ref{sec:theory} introduces the redshift weights and BAO modeling for the correlation functions. \S~\ref{sec:datasets} describes the simulations and datasets used in this work. 
In \S~\ref{sec:analysis}, we describe the redshift weighting algorithm 
in detail and provide the fitting model. We
present our DR14 data and mock results in \S~\ref{sec:results} and show the improvement due to redshift weighting. 
We share an outlook of the BAO constraints from the final quasar sample in \S~\ref{sec:outlook}. We emphasize the efficacy of redshift weighting for such a sample. 
We conclude in section \S~\ref{sec:discussion} with a discussion of our results.

\section{Theory}
\label{sec:theory}
\subsection{Distance Redshift Relation}
\label{sec:distance_redshift}

Following \citet{Zhu15}, we parameterize the distance-redshift relation, relative to a fiducial cosmology, as a Taylor series. Denoting the comoving radial distance by $\chi(z)$, we have 
\begin{equation}
\label{eq:chi_ratio}
	\frac{\chi\left(z\right)}{\chi_{f}\left(z\right)}=\alpha_{0}\left(1+\alpha_{1}x+ \frac{1}{2}\alpha_{2}x^{2} + \cdots \right).
\end{equation}
In the above parametrization, $\chi_f(z)$ labels the fiducial comoving radial distance and $x(z) \equiv\chi_f(z)/\chi_f(z_0) - 1$. Here $z_0$ is a pivot redshift chosen at convenience within the redshift range of the survey. 

The ratio between the fiducial and true Hubble parameter $H = 1/\chi'(z)$ is given by
\begin{equation}
\label{eq:h_ratio}
\frac{H_f(z)}{H(z)} = \alpha_0 \left[1+\alpha_1 + (2\alpha_1 + \alpha_2) x + \frac{3}{2}\alpha_2 x^2  + \cdots \right].
\end{equation}

Once the parameters $\alpha_0, \alpha_1$, and $\alpha_2$ are inferred from the sample, it is straightforward to reconstruct the measured distance-redshift relation and Hubble parameter from our expansion. When the fiducial cosmology coincides with
the true cosmology, one will measure $\alpha_0 = 1$, $\alpha_1 = 0$, and $\alpha_2 = 0$. 

We can easily extend this Taylor series to higher orders, but the parametrization to the first order can recover the
distance-redshift relation to sub-percent levels across a wide
range of redshifts and cosmologies. Even for the rather extreme $\Omega_M = 0.2$ and $\Omega_M = 0.4$ cases, the errors are less than 0.3 per cent over the redshift range of the eBOSS DR14 quasar sample $0.8 < z < 2.2$. 
We will thus focus on $\alpha_0$ and $\alpha_1$ and drop all higher order terms in the BAO analysis presented in this paper.  

A simple relation exists between our parametrization and the $(\alpha, \epsilon)$ or $(\alpha_\perp, \alpha_\parallel)$ parametrization \citep{Padmanabhan08Ani, Xu2013} used in recent BAO analyses \citep{AndersonDR11, DR12}. 
In these analyses, the deformation of the separation vectors between pairs of galaxies 
are parameterized by an `isotropic dilation' parameter $\alpha(z)$ and an `anisotropic warping' parameter $\epsilon(z)$. 
In the plane parallel limit, $\alpha$ and $\epsilon$ are related to the comoving distance and Hubble parameter by 
\begin{align}
\label{eq:pp_alpha}
	\alpha(z) &= \left[\frac{H_f(z)\chi^2(z)}{H(z)\chi_f^2(z)}\right]^{1/3}  \\
\label{eq:pp_epsilon}
	\epsilon(z) &= \left[\frac{H_f(z)\chi_f(z)}{H(z)\chi(z)}\right]^{1/3} - 1.
\end{align}
Together with Eq.~\ref{eq:chi_ratio} and Eq.~\ref{eq:h_ratio}, 
we can relate $\alpha(z)$ and $\epsilon(z)$ to $(\alpha_0,\alpha_1)$. Working to linear order in
$\alpha_1$, we have
\begin{align}
\label{eq:alpha}
	\alpha(z) &=  \alpha_0\left(1+\frac{1}{3}\alpha_1+\frac{4}{3}\alpha_1x\right) \\
\label{eq:eps}
	\epsilon(z) &= \frac{1}{3}\alpha_1 + \frac{1}{3}\alpha_1x.
\end{align}

\subsection{Redshift-weighted Correlation Function}
\label{sec:weighted estimators}

Modeled on \citet{TTH} as an extension of \citet{FKP}, \citet{Zhu15} developed the general formalism for a set of redshift weights for BAO analyses. The weights optimize the measurement of the parameters $\alpha_0$ and $\alpha_1$ in our distance-redshift relation parametrization. These weights can be expressed as the product 
of two components as $d{\cal W}w_{\ell, i}$. The first component is the commonly used FKP weights in galaxy surveys
\begin{equation}
	d{\cal W}(z) = \left(\frac{\bar n}{{\bar n} P + 1}\right)^2 dV(z).
\end{equation}
This expression corresponds to the inverse covariance of the power spectrum in redshift slices. 

The second component $w_{\ell, i}$ is a linear combination of 1 and $x$. The specific linear combination depends on the parameter ($\alpha_0$ or $\alpha_1$, indicated by the subscript $i$) and the multipoles (monopoles or quadrupoles, indicated by $\ell$) in question.  
The redshift weights are generalizations of the FKP weights produced by up-weighting the regions where the signal is most sensitive to the model parameters, in addition to balancing the quasars by number densities. 

Since the additional weights $w_{\ell, i}$ are linear combinations of 1 and $x$, it is convenient to calculate
correlation functions weighted by 1 and $x$ instead of the original weights. 
We construct the `1-weighted' and `$x$-weighted' correlation functions as
\begin{align}
	\xi_{\ell,1}(r) & =  \frac{1}{N} \int  d\mathcal{W}(z) \xi_{\ell,g}(r, z)\\
	\xi_{\ell,x}(r) & =  \frac{1}{N}\int d\mathcal{W}(z)x(z) \xi_{\ell,g}(r, z)
\end{align}
where $N=\int d\mathcal{W}$ is a convenient choice of normalization and where the correlation function $\xi_{\ell, g} = b^2 \xi_{\ell,m}$, where $b$ is the linear bias.

In these models, the integrals are over the redshift range of the sample. They can be efficiently computed as summations over contributions from discrete redshift slices. We follow the same procedure as in \citet{Zhu16} and 
calculate contributions from redshift slices of width $\Delta z = 0.1$ within the redshift range [0.8, 2.2]. 
In each redshift slice, given $\alpha_0$ and $\alpha_1$, we compute the `isotropic dilation' parameter $\alpha(z)$ and `anisotropic warping' parameter $\epsilon(z)$ according to
Equation~\ref{eq:alpha} and Equation~\ref{eq:eps} at different redshifts.
This feature is distinct from traditional analyses in which $\alpha$ and $\epsilon$ are only measured at the 
`effective' redshift of the sample. We will describe how $\alpha$ and $\epsilon$ shift and distort the correlation function in Sec.~\ref{sec:mis-estimate}.
Thus, our model parameters $\alpha_0$ and $\alpha_1$, which we will obtain directly from our fits to the measured $\xi$, provide constraints on $\alpha(z)$ and $\epsilon(z)$ given our perturbative model


\subsection{Fitting the Correlation Function}

We fit the correlation function
with the ESW template given in \citet{ESW}\footnote{Also see \citet{WhiteZeldovich} and \citet{Vlah2016} for a more advanced perturbation theory based template. }. 
We will outline the ESW template below and explain its ingredients and how mis-estimate of the cosmology 
distorts the correlation function and how to model it. 
The fitting model is similar as in recent BOSS BAO analyses \citep{AndersonDR11, DR12}. 

\subsubsection{BAO modeling}

Our template combines the supercluster infall of linear theory \citep{Kaiser1987} and the Finger-of-God (FoG) effect from non-linear growth of structure.

In Fourier space, we use the following 2D non-linear power spectrum template 
\begin{equation}
	P_t(k,\mu) = (1+\beta \mu^2)^2 F(k,\mu, \Sigma_s) P_{\rm dw}(k,\mu).
\end{equation}
The $(1+\beta\mu^2)^2$ term describes the Kaiser effect \citep{Kaiser1987} - distortion caused by coherent infall of objects towards the cluster center. Here $\beta = f/b$ where $f$ is the cosmological growth rate of structure and $b$ is the large scale bias. The $F(k,\mu,\Sigma_s)$ factor represents the Finger-of-god (FoG) effect - elongation in the observed structure along the line-of-sight direction given rise by large random velocities in inner virialized clusters. We model 
the FoG factor \citep{Park1994, Peacock&Dodds1994} as 
\begin{equation}
	F(k, \mu, \Sigma_s) = \frac{1}{1+k^2\mu^2 \Sigma_s^2}
\end{equation}
where $\Sigma_s$ denotes the streaming parameter to account for the dispersion due to random peculiar velocities within clusters. See \citet{White2015} for a comprehensive discussion of various streaming models. 

The `de-wiggled' power spectrum $P_{\rm dw}$ in the template takes the form 
\begin{equation}
	P_{\rm dw}(k,\mu) = [P_{\rm lin}(k) - P_{\rm nw}(k)] \exp\left[-\frac{k_\parallel^2 \Sigma_\parallel^2 + k_\perp^2 \Sigma_\perp^2}{2}\right] + P_{\rm nw}(k).
\end{equation}
In the equation above, $P_{\rm lin}(k)$ is the power spectrum from CAMB \citep{CAMB}. The no-wiggle power spectrum
$P_{\rm nw}(k)$
is the smoothed power spectrum \citep{Eisenstein&Hu} that removes the baryonic wiggles.
In the de-wiggled power spectrum template, the Gaussian damping term models the degradation of the BAO due to non-linear structure growth. Redshift space distortions make this damping anisotropic, which is captured by the difference in the parallel and perpendicular streaming scales $\Sigma_\parallel$ and $\Sigma_\perp$ along and across the line-of-sight. In our analyses, we fix $\Sigma_\perp = 3 h^{-1} {\rm Mpc}$ and $\Sigma_\parallel = 6 h^{-1}{\rm Mpc}$. These values are based on estimates of the streaming parameters \citep{Crocce&Scoccimarro2006, Crocce&Scoccimarro2008, Matsubara2008} at median redshift of the sample $z = 1.5$. We also vary these parameters and find the fitting result to be insensitive to these choices.

The 2D power spectrum template can be decomposed into multipole moments as
\begin{equation}
	P_{\ell, t} = \frac{2 \ell +1}{2}\int_{-1}^{1} P_t(k,\mu) L_\ell(\mu) d\mu .
\end{equation}
where $L_\ell$ is the Legendre polynomial. The correlation function multipoles and power spectrum multipoles are Fourier transform pairs and can be obtained as
\begin{equation}
	\xi_{\ell, t} = i^\ell \int \frac{k^3 d\log k}{2\pi^2} P_{\ell, t}(k) j_\ell(kr).
\end{equation}

\subsubsection{Modeling the mis-estimate of cosmology}
\label{sec:mis-estimate}
The difference between the true and fiducial cosmology distorts the calculated correlation function. 
We review how the distorted correlation function can be modeled in terms of the `isotropic dilation' and `anisotropic warping' parameters $\alpha$ and $\epsilon$ \citep{Padmanabhan08Ani, Xu2013}.
The approach here is the same as in Sec 2.2 of \citet{Zhu16} and we refer the readers to that paper for details. 
In summary, the `true' quasar separation and the cosine of the angle between the separation vector and line-of-sight 
are expressed in terms of the fiducial values by 
\begin{align}
\label{eq:true_r}
	r      &= \alpha r^{\rm f}\sqrt{(1+\epsilon)^4 (\mu^{\rm f})^2 + (1+\epsilon)^{-2}[1-(\mu^{\rm f})^2]} \\
	\mu &= \cos[\arctan[(1+\epsilon)^{-3}\tan(\arccos \mu^{\rm f})]].
\label{eq:true_mu}
\end{align}

Given $\alpha_0$ and $\alpha_1$, we can calculate $\alpha(z)$ and $\epsilon(z)$ at all redshifts within the redshift range of the sample. These $\alpha$ and $\epsilon$ indicate how $r$ and $\mu$ are distorted at all redshifts, allowing us to incorporate the mis-estimate of the cosmology into model correlation functions.

\section{DATASETS}
\label{sec:datasets}

\subsection{SDSS DR14 Quasar Sample}

The observational dataset is the eBOSS \citep{Dawson2016} quasar sample released as part of the SDSS-IV \citep{Blanton2017}. 
The survey has an effective area of 1192 deg$^2$ in the Northern Galactic Cap (NGC)
and 857 deg$^2$ in the Southern Galactic Cap (SGC). The quasar target selection is presented in 
\citet{Ross2012} and \citet{Myers2015}. Quasars selected that do not have a known redshift are selected for spectroscopic observation. Spectroscopy is obtained through the BOSS double-armed spectrographs \citep{Smee2013}. 
In our DR14 sample, we applied veto masks as in \citet{Reid2016}. To correct for missing targets, redshift failures, fiber collisions, depth dependency, and Galactic extinction, we utilize completion weights and systematic weights as in \citet{Laurent2017} and \citet{Ross2017}.

\subsection{Simulations}

We test our algorithm
on mock catalogues created by using the `quick particle mesh' (QPM)
method \citep{WhiteQPM}. These mock catalogues are constructed 
to simulate the clustering and noise level of the eBOSS DR14 quasar sample. The mock catalogues are based on 100 low force- and mass-resolution
particle-mesh N-body simulations. Each uses $2560^3$ particles in a box of side length $5120 h^{-1}{\rm Mpc}$. 
The simulations assume a flat $\Lambda$CDM cosmology, with $\Omega_m = 0.31$, $\Omega_b h^2 = 0.0220$, $h = 0.676$, $n_s = 0.97$, and $\sigma_8 = 0.8$. 
Each simulation starts at $z = 25$ using 
second order Lagrangian perturbation theory. 
The catalogues span the redshift range of $z=0.8$ to $2.2$ and cover both the northern and southern Galactic cap of
the eBOSS footprint. 
The halo occupation of quasars is parameterized according to the five-parameter halo occupation distirbution (HOD) presented in \citet{Tinker2012}. 

Rotating the orientations of the 100 simulated cubic boxes, we identify four configurations with less than 1.5 per cent overlap. This enables us to produce 400 QPM mocks for both Galactic caps. Veto masks are applied in the same way as for the data. FKP weights \citep{FKP} are applied assuming $P_0 = 6000 h^{-3}{\rm Mpc}^3$. Redshift smearing is applied according to \citet{Dawson2016}. For specifics of these eBOSS quasar mocks, we refer the readers to \citet{DR14alphabetical}. 

\section{Analysis}
\label{sec:analysis}

\subsection{Computing the weighted correlation functions}

We analyze the simulations in a manner similar to previous BOSS analyses \citep{AndersonDR11, DR12}. 
We refer the reader to those papers for more detailed descriptions. 
We do not apply density field reconstruction \citep{EisensteinRecon07}, 
as it is not expected to be efficient or significant for this sample due to the low density of quasars.

To compute the weighted correlation functions from the catalogues, 
we use a modified version of the Landy-Szalay estimator \citep{LS}.  
In addition to weighting each quasar/random by the FKP weight,
we also weight each pair of quasars/randoms by $x$ to construct the $x$-weighted correlation
functions. 
Since the redshift separation between a pair that contributes to the correlation function is small, 
we simply use the mean redshift of each pair to compute $x$. 
The weighted 2D correlation functions are given by
\begin{equation}
\label{eq:weighted_LS}
	\xi_w^{\rm data}(r,\mu)=\frac{\widetilde{DD}(r,\mu) - 2 \widetilde{DR}(r,\mu) + \widetilde{RR}(r,\mu)}{RR(r,\mu)}
\end{equation}
where $\widetilde{DD}$, $\widetilde{DR}$ and $\widetilde{RR}$ include the additional pair weight, 
whereas $RR$ in the denominator does not. 
From the 2D correlation function, one can compute the monopole and quadrupole moments as
\begin{equation}
	\xi_{\ell,w}^{\rm data}(r) = \frac{2\ell + 1}{2} \int_{-1}^1 \xi_w^{\rm data}(r,\mu) L_\ell(\mu) d\mu.
\label{eq:xi_data}
\end{equation}

We consider two cases: an unweighted sample using only the FKP weight and a weighted sample uses both the `1' and `$x$' weights. For both cases, we treat the quasar sample as a unified one without splitting it into redshift bins. 

\subsubsection{The Fiducial Fitting Model}

We fit our correlation functions to
\begin{equation}
	\xi^{\rm fit}_{\ell,w}(r) = B_w^2 \xi_{\ell,w}(r) + A_{\ell,w}(r)
\end{equation}
where $\xi_{\ell,w}(r)$ is the weighted correlation function and 
$A(r)$ absorbs un-modeled broadband
features including redshift-space distortions and scale-dependent bias
following \citet{AndersonDR11}. We assume
\begin{equation}
A_{\ell,w}\left(r\right)=\frac{a_{\ell,w,1}}{r}+a_{\ell,w,2}.
\end{equation}
We allow a multiplicative factor $B_w^2\sim1$ to vary in order to adjust the amplitudes of the correlation functions. 
The quantities $B_w^2$ determines the amplitudes of the monopole and quadrupole together, while $\beta$ sets the relative 
amplitude between the two. 

In our fiducial weighted fits, we use a total of 13 fitting parameters : $\alpha_0,\alpha_1,\beta, B_1, B_x$, and 8 nuisance parameters to absorb the broadband features.  We use the fiducial fitting range $ 48 < r < 184 {h^{-1} \rm Mpc}$ with 8 ${h^{-1}\rm Mpc}$ bins. 

%
%
%

\begin{figure*}
\centering 
\includegraphics[width=0.9\textwidth]{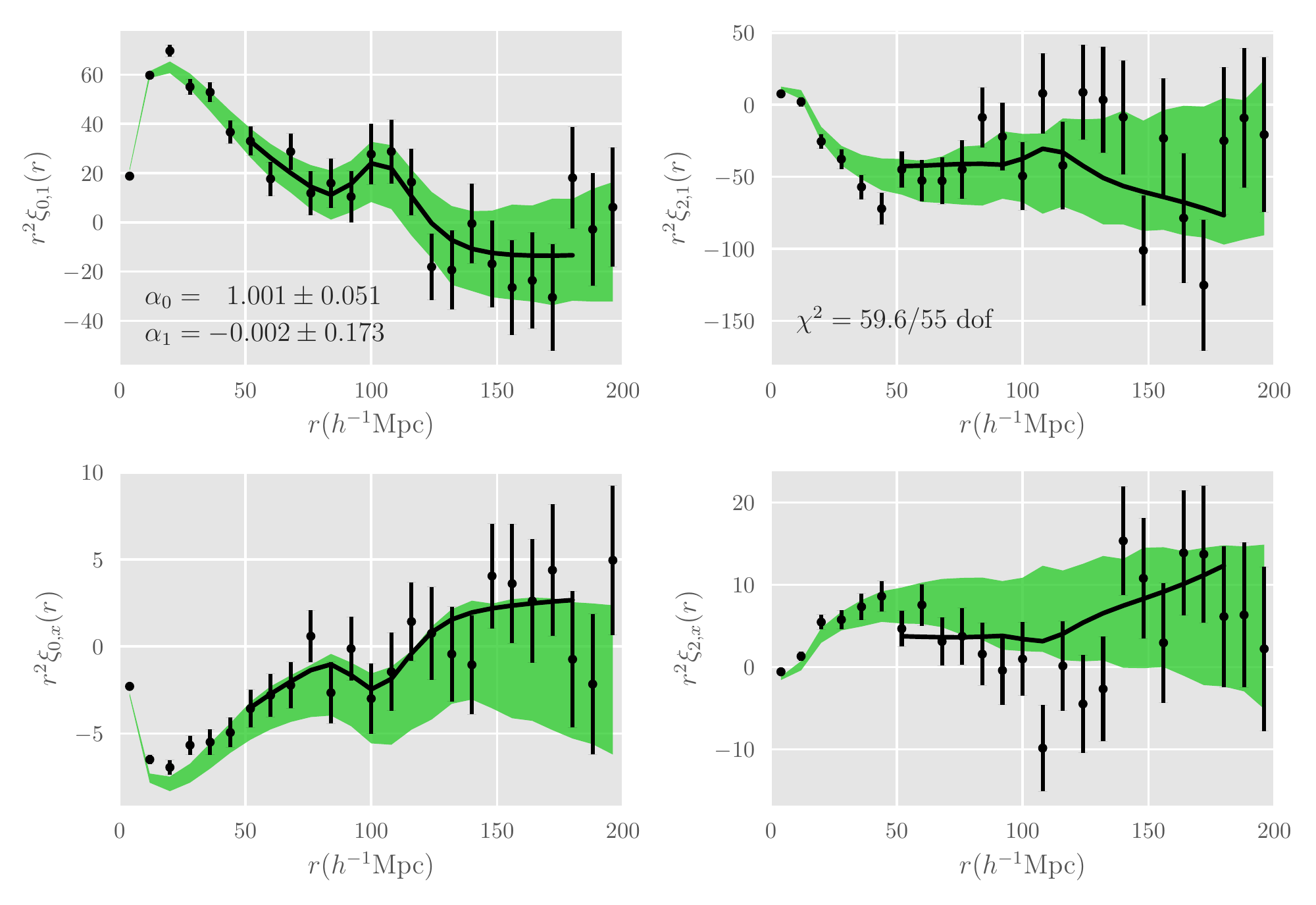}
\caption{The DR14 quasar correlation function and the average QPM mock correlation functions. The black circles with error bars are the correlation function multipoles from the DR14 sample. The top panels display the `unweighted' monopoles (left) and quadrupoles (right), while the bottom pair show the `$x$-weighted' ones. The associated error bars are $1\sigma$ errors of the mocks. The solid black line passing through the black points show the best-fit to the DR14 points with relevant statistics on the top panels. 
The green bands in each on the figure represent the average monopoles (left) and quadrupoles (right) from the 400 QPM mocks with 1 standard deviation errors.  
The error bands plotted are that of an individual mock, which are $\sqrt{400}$ times larger than that of the 
average correlation function.  
The `$x$-weighted' monopoles and quadrupoles are inverted as compared to the `unweighted' ones, due to an overall negative weight. 
 }
\label{fig:average}
\end{figure*}

\subsection{Parameter Inference}

We assume the likelihood function is a multi-variate Gaussian. The posterior distribution of $\alpha_0$ and $\alpha_1$ can be written as 
\begin{equation}
	p(\alpha_0, \alpha_1) \propto e^{-\chi^2(\alpha_0, \alpha_1)/2}
\end{equation}
where $\chi^2$ is given by 
\begin{equation}
	\chi^2 = {\mathbfss D}  {\mathbfss C}^{-1} {\mathbfss D}^T, 
\end{equation}
where ${\mathbfss C}$ represents the covariance matrix and ${\mathbfss D}$ is the difference between the data and model vectors. We calculate ${\mathbfss C}$ as the sample covariance matrix from the mocks and apply the correction factor defined in \citet{Hartlap07} and \citet{PercivalCovar}. 

Given $\alpha_0$ and $\alpha_1$, we minimize the $\chi^2$ through a simplex algorithm \citep{nelder1965simplex} designed to handle the non-linear parameters, while
the linear nuisance parameters are obtained using a least-squares method nested within the simplex. 
The simplex algorithm searches the non-linear parameter space until the best-fit parameters 
that minimize $\chi^2$ are achieved.

We calculate the likelihood surface through computing best-fit $\chi^2$ on a two-dimensional grid for $ 0.7< \alpha_0 < 1.3$ and $ -0.5 < \alpha_1 < 0.5$ at spacings of 0.01 and 0.02, respectively. 
The likelihood surface enables one to calculate the distribution of $\alpha_0$ and $\alpha_1$.
The low signal-to-noise BAO feature of some mocks causes the nuisance polynomial to 
dominate the model correlation function. To avoid these undesirable cases, 
we adopt a Gaussian prior for $\beta$ centered around 0.4 with width 0.2. 
We also adopt a Gaussian prior on $B_1^2$ and $B_x^2$ at 1 with width 0.2. 
To suppress the unphysical downturns in $\chi^2$, we have applied Gaussian priors of width 0.1 centered around $\alpha_0= 1$ and width 0.2 centered around $\alpha_1=0$. These priors do not dominate our calculation of the likelihood of $\alpha_0$ and $\alpha_1$. Their implications are discussed in more detail in Sec.~\ref{sec:results}.
 
%

\section{Results}
\label{sec:results}

The fits to the mock correlation functions assume the QPM 
cosmology as the fiducial cosmology using a pivot redshift $z_0 = 1.8$. The fitting procedure and the model are outlined in \S~\ref{sec:analysis}. 


Fig.~\ref{fig:average} shows the DR14 quasar correlation functions and the average of these from 400 mocks. 
The DR14 quasar correlation functions are indicated as points with error bars. 
The bands in the figure correspond to the $1 \sigma$ error for individual mocks. 
The mocks are consistent with the DR14 points. The quadrupole moments show significant noise. 
Despite the uncertainties, the monopole moments demonstrate a clearly visible acoustic feature in both the `1-weighted' and `$x$-weighted monopoles. 

\begin{figure*}
\centering
\includegraphics[width=\textwidth]{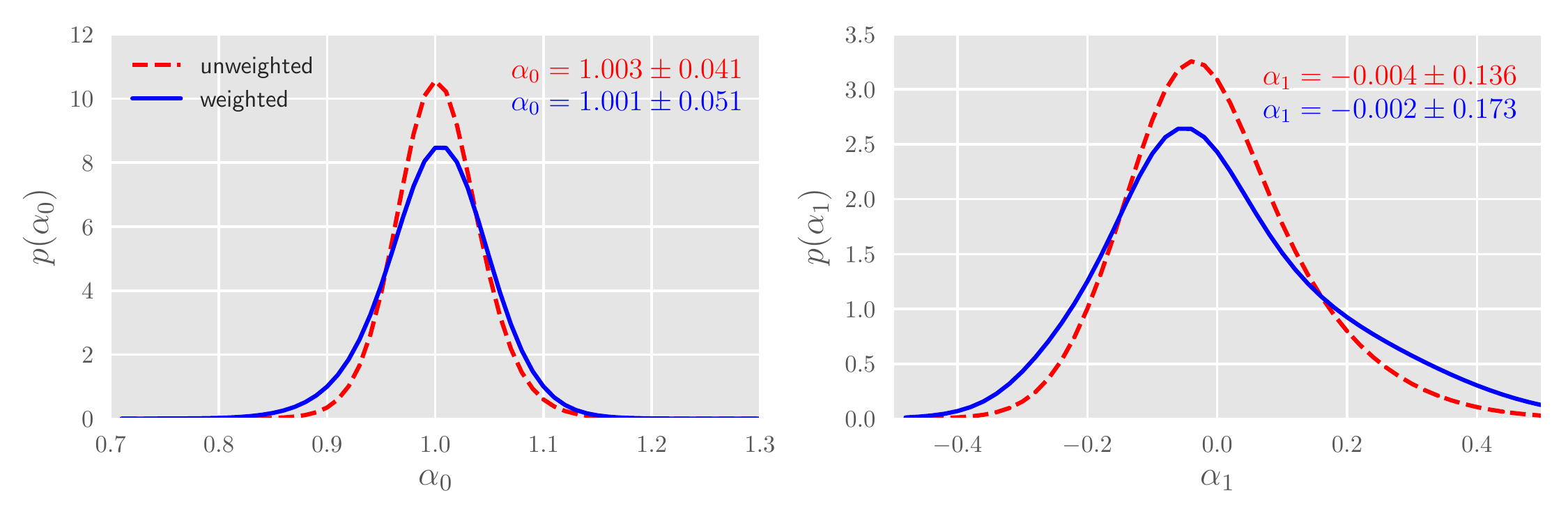}
\caption{Distribution of $\alpha_0$ and $\alpha_1$ from the DR14 fits. The left and right panels show the derived $p(\alpha_0)$ and $p(\alpha_1)$ distributions respectively. The red dashed lines correspond to the distribution from fitting the `unweighted' estimator; the blue solid lines correspond to the sharpened distribution we obtained from applying redshift weighting. The mean and standard deviation of both distributions are labeled in the panels. }
\label{fig:likelihood}
\end{figure*}

The thick black line are the best-fit to the DR14 data points with relevant statistics labeled on the figure. 
In the fiducial case, we measure $\alpha_0 = 1.001\pm 0.051$ and $\alpha_1 = -0.002\pm 0.173$. 
The `unweighted' fits without redshift weighting yield $\alpha_0 = 1.003\pm0.041$ and $\alpha_1 = -0.004\pm0.136$. The distribution of $\alpha_0$ and $\alpha_1$ measured from the DR14 quasar sample is shown in Fig.~\ref{fig:likelihood}. For the DR14 sample, applying redshift weighting does not yield reduction in the size of the error bars for the measured $\alpha_0$ and $\alpha_1$. 


\begin{table}
\caption{
BAO fitting results of the DR14 quasar data and QPM mocks. Our fiducial analysis assumes a pivot redshift of $z_0 = 1.8$ and a fitting range of $48 < r < 184h^{-1}{\rm Mpc}$ with $8h^{-1}{\rm Mpc}$ binning. The fiducial analysis utilises redshift weighting. The mock results shown here are the inverse variance weighted average of the 400 QPM mock fits.
}
\label{tab:bestfit_tab}
\begin{tabular}{lcr}
    \hline \hline
    \multicolumn{1}{l}{Model} &  $\alpha_0$ & \multicolumn{1}{c}{$\alpha_1$} \\ \hline  
     & DR14 Results &\\ \hline 
      Fiducial		       			            	& $1.001\pm 0.051$  & $0.002 \pm 0.173$  \\  
      Fiducial, unweighted               			& $1.003\pm 0.041$  & $-0.004\pm 0.136$   \\ 
       Fit w/ $\Sigma_s = 2 \ h^{-1}{\rm Mpc}$ 		& $1.004\pm 0.052$   & $0.014\pm 0.172$ \\         
       $(\Sigma_\perp,\Sigma_\parallel) = (4, 8)\ h^{-1}{\rm Mpc}$ & $1.002\pm 0.051$ & $0.007\pm 0.172$ \\ 
       Fi w/ poly3                                                    & $1.001\pm 0.048$ & $-0.023\pm 0.175$ \\
       Fit w/o $x$-weighted quadrupole                 & $1.006\pm 0.043$ & $0.013\pm 0.134$ \\
       $\alpha_0$ only				              &  $0.996\pm 0.031$   & \multicolumn{1}{c}{$-$}  \\
       $48 < r < 136 \ h^{-1} {\rm Mpc} $                     & $0.999\pm 0.053$  & $-0.015\pm 0.167$ \\
       $48 < r < 160 \ h^{-1}{\rm Mpc} $ 			& $0.987\pm 0.061$  & $-0.009\pm 0.193$  \\ 
       $\Delta r = 4 \ h^{-1}{\rm Mpc}$   			& $0.997\pm 0.049$   & $0.090\pm 0.165$ \\  
       $z_{\rm pivot} = 1.2$					 & $1.002\pm 0.072$ & $-0.002\pm 0.131$   \\
       $z_{\rm pivot} = 2$					& $0.999\pm 0.049$ & $0.001\pm 0.179$ \\
        \hline
       & Mock Results &   \\
       \hline
       Fiducial 		          				& $0.992\pm 0.052$ & $0.001\pm 0.141$  \\ 
       Fiducial, unweighted      				 & $0.998\pm 0.054$ & $0.014\pm 0.157$  \\  
       Fit w/ $\Sigma_s = 2 \ h^{-1}{\rm Mpc}$ 		& $0.993\pm0.054$   & $0.003\pm0.144$ \\   
       $(\Sigma_\perp,\Sigma_\parallel) = (4, 8)\ h^{-1}{\rm Mpc}$ & $0.992\pm0.052$ & $0.003\pm0.141$ \\ 
       Fit w/ poly3                                                   & $0.991\pm 0.053$ & $0.001\pm 0.147$ \\
       Fit w/o $x$-weighted quadrupole                 & $0.993\pm 0.052$ & $0.001\pm 0.143$ \\
       $48 < r < 136 \ h^{-1}{\rm Mpc} $                     & $0.988\pm 0.055$  & $-0.006\pm 0.143$ \\
       $z_{\rm pivot} = 1.2$ 					& $0.991\pm0.067$  & $-0.014\pm0.115$\\
       $z_{\rm pivot} = 2$ 					& $0.993\pm 0.050$ & $-0.001\pm 0.146$\\
       \hline
       & `4x' Mock Results & \\
       \hline
       `4x' mocks, Fiducial					&  $0.995\pm 0.028$  & $0.001\pm0.077$   \\  
       `4x' mocks, unweighted  				&  $0.996\pm 0.031$ &$0.017\pm0.105$  \\ 
       `4x' mocks, $z_{\rm pivot} = 1.2$ 		&  $0.993\pm0.040$ & $-0.001\pm0.060$ \\
       `4x' mocks, $z_{\rm pivot} = 2$			&  $0.996\pm0.026$ & $-0.001\pm0.081$ \\
       \hline \hline
\end{tabular}
\end{table}

We test the robustness of our result by varying various aspects of the fit including the fitting range, binning,  streaming parameters, and pivot redshift. The results all agree within $1\sigma$ uncertainties. Table~\ref{tab:bestfit_tab} presents a summary of our fitting results. 
In the table, $poly3$ corresponds to fitting with a third degree nuisance polynomial of the form $A(r) = a_1/r^2 + a_2 / r + a_3$. 
In addition, we perform an isotropic BAO fit by setting $\alpha_1 = 0$ and only allowing $\alpha_0$ to vary. 
This analysis produces $\alpha_0 = 0.996\pm0.031$, consistent with the result $0.994\pm0.037$ in \citet{DR14alphabetical}. The small discrepancy in the error could be due to differences in the applied priors, as  \citet{DR14alphabetical} restricts to 
the prior range $0.8 < \alpha < 1.2$. In our calculation of the likelihood, we use a larger prior range $0.7 < \alpha_0 < 1.3$ and a Gaussian prior of width 0.1 centered around $\alpha_0=1$. 

\begin{figure}
\includegraphics[width=0.48\textwidth]{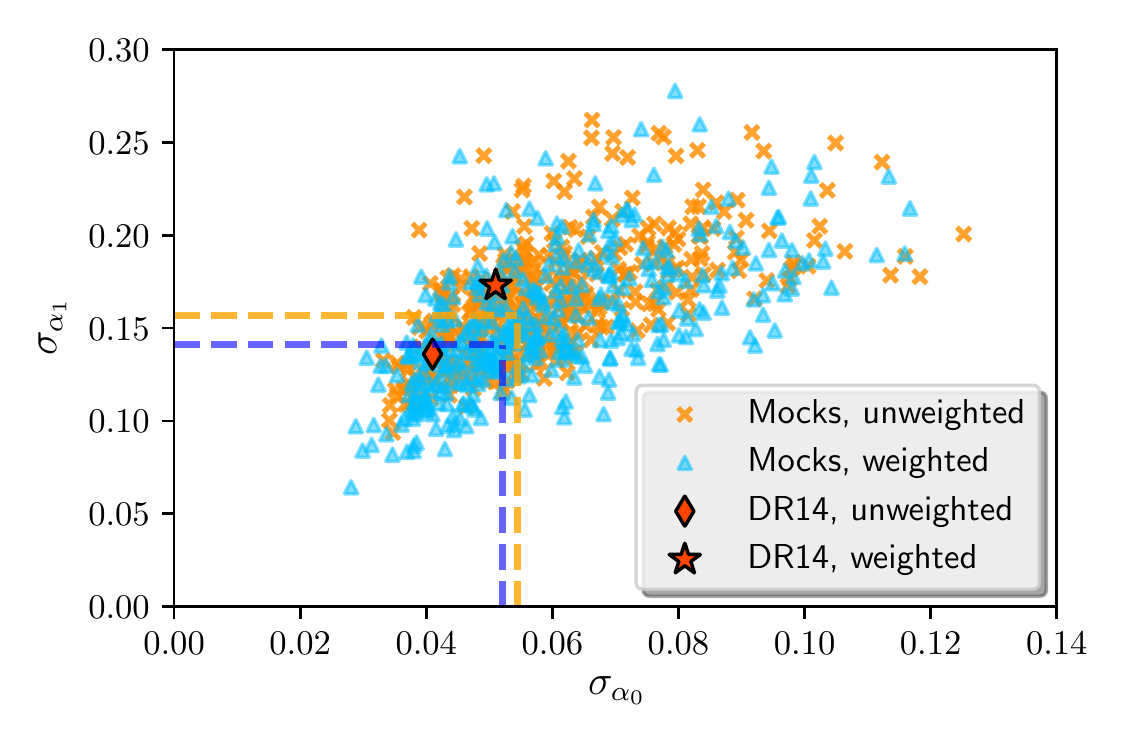}
\caption{The $\sigma_{\alpha_0}$ and $\sigma_{\alpha_1}$ values measured from the 400 mocks and from the DR14 sample. The blue triangles correspond to the `weighted' measurement errors and the orange triangles are the `unweighted' values. The errors denoted by the horizontal and vertical dashed lines are the errors of the inverse variance weighted average of the mock results, multiplied by $\sqrt{400}$ for ease of comparison with individual mock points. Our DR14 $\sigma_{\alpha_0}$ and $\sigma_{\alpha_1}$ are labeled as the red star (`weighted') and diamond (`unweighted'). The DR14 point falls within the locus of mock points. }
\label{fig:std}
\end{figure}

To validate our methodology, we fit 400 QPM mocks and measure $\alpha_0$ and $\alpha_1$.  
Our fiducial cosmology is the same as the simulation cosmology. 
Therefore, we expect our measurements to agree with $\langle \alpha_0\rangle = 1$ and $\langle \alpha_1 \rangle = 0$ within uncertainty if the measurements are unbiased. A summary of the mock results can be found in Table~\ref{tab:bestfit_tab}. 
We indeed verify our method to yield unbiased estimators of $\alpha_0$ and $\alpha_1$.

The errors of $\alpha_0$ and $\alpha_1$ measured from the 400 QPM mocks are indicated as blue points in Fig.~\ref{fig:std}. 
The orange points in the background show the errors from the `unweighted' fits. 
The fitted DR14 data point is also displayed. The mock $\alpha_0$ and $\alpha_1$ errors are representative of the DR14 errors. 

We compare the $\sigma_{\alpha_0}$ and $\sigma_{\alpha_1}$ obtained from the `unweighted' and `weighted' analysis mock by mock. Among the 400 mock measurements, 221 produce an improved $\sigma_{\alpha_0}$, and 275 show an improved $\sigma_{\alpha_1}$ when we apply the redshift weights. 
These values correspond to 55 per cent and 69 per cent of the mocks. 
Given the magnitude of these percentages, it is not surprising that redshift weighting does not yield smaller $\sigma_{\alpha_0}$ and $\sigma_{\alpha_1}$ errors for the current DR14 sample. 

Overall, however, redshift weighting does shrink the measured error bars. 
We aggregate the mock measurements of $\alpha_0$ and $\alpha_1$ through inverse variance weighting to minimize the variance of the weighted average. 
Each mock measurement of $\alpha_0$ and $\alpha_1$ is weighted in inverse proportion to its variance. 
We obtain this weighted average as ${\hat \alpha} = \frac{\sum \alpha_i/\sigma_{\alpha_i}^2}{\sum 1/\sigma_{\alpha_i}^2}$. 
The summation is performed over the 400 mocks. 
The error of ${\hat \alpha}$ is given by $\sigma({\hat \alpha}) = 1/\sqrt{\sum \frac{1}{\sigma_{\alpha_i}^2}}$.  
This error is scaled by $\sqrt{400}$ for ease of comparison with errors from individual mock measurements.
The aggregated mock statistics are presented in Table~\ref{tab:bestfit_tab}. 
We observe a decrease in $\sigma_{\alpha_1}$ from 0.157 without redshift weights to 0.141 with redshift weights. 
This change corresponds to a 10 per cent decrease. 
We will further comment on the magnitude of this improvement in \S~\ref{sec:outlook}.  


The joint likelihood distribution of $\alpha_0$ and $\alpha_1$ allows us an estimate to be made of the joint distribution of $\chi$ and $H$. 
To perform this calculation, we first draw random variables from the joint distribution of $\alpha_0$ and $\alpha_1$. 
We reconstruct the distance-redshift relation $\chi(z)$ and Hubble parameter $H(z)$ from Eq.~\ref{eq:chi_ratio} and Eq.~\ref{eq:h_ratio} with the drawn $\alpha_0$ and $\alpha_1$. 
This approach enables us to obtain an estimated joint distribution of $\chi$ and $H$.  
It is then straightforward to calculate statistics of $\chi$ and $H$. 
Since these $\chi$ and $H$ measurements at different redshifts are derived from the same model of the distance-redshift relation, they are highly correlated. 
To use our result for cosmological comparisons, it is advisable to directly use the joint likelihood distribution of $\alpha_0$ and $\alpha_1$ we measured.


Our parametrization of the distance-redshift relation and Hubble parameter allows one to obtain constraints for both at all redshifts within the range of the sample. 
In Table~\ref{tab:constraints} we produce $D_M$ and $H$ measurements at several redshifts. 
We also derive spherically-averaged distance measurement $D_V$ from our $D_M$ and $H$ measurements. 
The measurements at these redshifts are highly correlated. 
We thus report the correlation matrix for $D_M$ and $H$ at only two redshifts $z_1=1$ and $z_2 = 2$ below as 
\begin{equation}
   \mathbfss{C} =
	\begin{blockarray}{ccccl}
   D_M(z_1) & H(z_1) & D_M(z_2) & H(z_2) \\
	\begin{block}{(cccc)l}
   		1 &  0.25 & 0.72 & 0.66 & D_M(z_1)\\
   		   & 1 & -0.48 & 0.85 &H(z_1)\\
		   &    &  1 & 0.00 &D_M(z_2)\\
		   & & & 1 &H(z_2)\\
	\end{block}
  \end{blockarray}
\end{equation} 
The correlation between $D_M(z=1)$ and $D_M(z=2)$ is quite substantial, as is the correlation between $H$ at $z=1$ and $z=2$. 
However, at both redshifts, the correlation between $D_M$ and $H$ is low. 
This behavior is not necessarily the case for a different choice of $z_1$ and $z_2$. 
There is a tradeoff between the correlation of $D_M$ and $H$ at the same redshift and the correlation between $z_1$ and $z_2$. 



\begin{table}
\caption{Constraints on $D_M(r_{\rm d, fid}/r_{\rm d})$ and $H (r_{\rm d} / r_{\rm d, fid})$ measured from the DR14 quasar sample from our analysis with redshift weighting. Also listed are the derived spherically averaged distance measurements $D_V(r_{\rm d, fid}/r_{\rm d})$ from our $D_M$ and $H$ measurements. The measurements at different redshifts are correlated. }
\label{tab:constraints}
\begin{tabular}{ cccc }
    \hline
        \multirow{2}{*}{Redshift} & \multirow{2}{2cm}{\centering $D_M (r_{\rm d, fid} / r_{\rm d})$ [Mpc]}  & \multirow{2}{1.8cm}{\centering $H (r_{\rm d} / r_{\rm d, fid})$  [${\rm km}\ {\rm s}^{-1}{\rm Mpc}^{-1}$]} & \multirow{2}{2cm}{\centering $D_V(r_{\rm d, fid} / r_{\rm d})^*$ [Mpc]} \\
        \\
         \hline
         $0.8$ & $2876\pm 304$ & $106.9\pm 4.9$   & $2646\pm 205$\\ 
         $1.0$     & $3405\pm 305$ & $120.7\pm7.3$   &  $3065\pm 182$\\
        $1.5$  & $4491\pm 272$ & $161.4\pm30.9$  &  $3840\pm 182$\\ 
        $2.0$      & $5325\pm 249$ & $189.9\pm32.9$  &  $4356\pm 300$\\
         $2.2$ & $5606\pm 255$ & $232.5\pm54.6$ &  $4514\pm 359$  \\
   \hline
\end{tabular}
\end{table}

In analyzing the BAO from the  BOSS DR 12 galaxy mock catalogues, \citet{Zhu16} reported that the distance and Hubble parameter measurements are insensitive to the choice of pivot redshifts. Our mock measurements confirm this finding. 

At different pivot redshifts, a large error in $\alpha_0$ is usually compensated by a smaller error in $\alpha_1$, and vice versa. Table~\ref{tab:bestfit_tab} lists fitting results at 3 different pivot redshifts $z_0=1.2, 1.8$, and 2. Selecting $z_0=2$ yields the smallest $\sigma_{\alpha_0}$ but has the largest $\sigma_{\alpha_1}$. 
Conversely, $z_0 = 1.2$ yields the largest $\sigma_{\alpha_0}$ but has the smallest $\sigma_{\alpha_1}$. 
When reconstructing $D_M$ and $H$ constraints from $\alpha_0$ and $\alpha_1$, the error from the two parameters compensate one another and makes the distance and Hubble parameter constraints insensitive to the choice of the pivot redshift. 

\begin{figure*}
\centering
\includegraphics[width=0.95\textwidth]{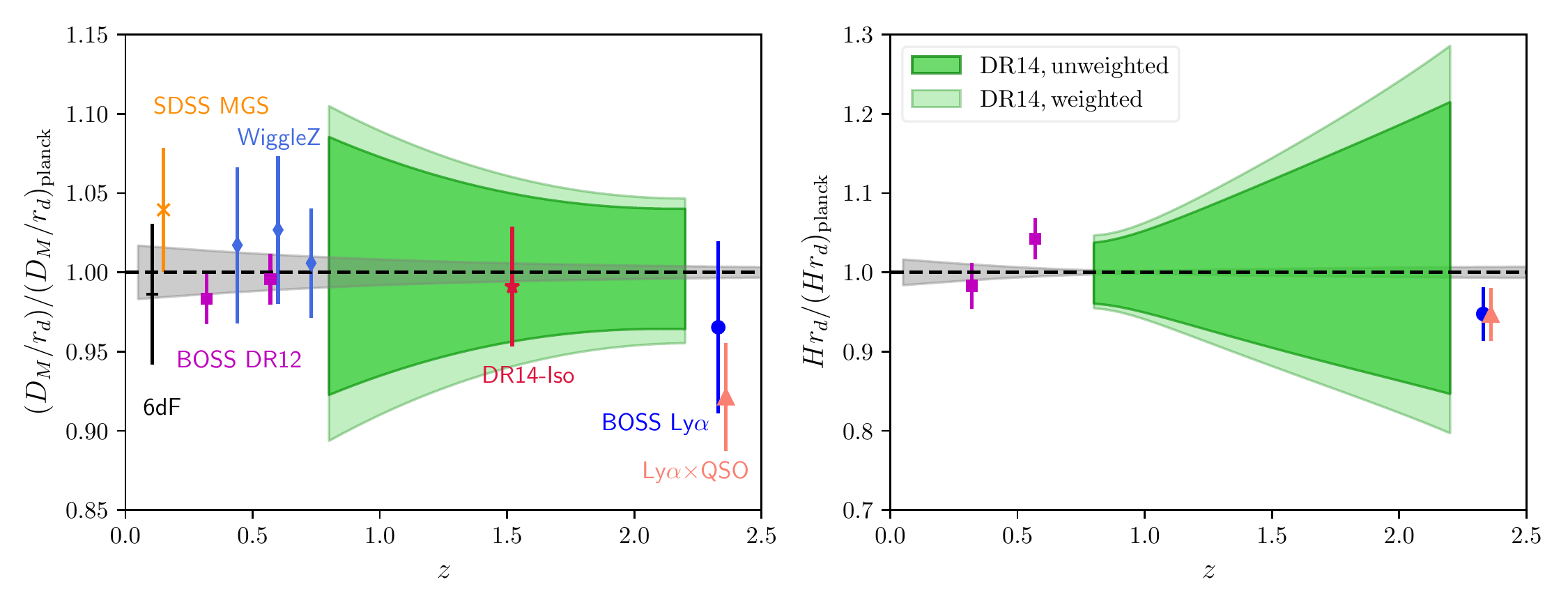}
\caption{Our DR14 $D_M$ and $H$ measurements from `unweighted' and 'weighted' analyses compared to the Planck flat-$\Lambda$CDM predictions. All error bands and error bars correspond to 1 standard deviation errors. Our DR14 measurements (green bands) are consistent with the The Planck results (grey bands) within uncertainty. We emphasize that the $D_M$ and $H$ measurements at different redshifts are highly correlated. We also show several recent measurements for comparison, some of which are spherically-averaged BAO distance measurements ($D_V$). See texts for descriptions of these additional measurements. }
\label{fig:planck_comparison}
\end{figure*}

We compare our results with recent measurements of $D_M$ and $H$. Fig.~\ref{fig:planck_comparison} displays our $D_M$ and $H$ measurements along with the $\Lambda$CDM prediction from Planck \citep{Planck2015}. 
Our distance and Hubble parameter measurements are in agreement with the Planck results within the $1\sigma$ uncertainty.  
We also show similar measurements in the literature: the BOSS DR12 results from \citet{DR12}, the BOSS Ly$\alpha$ from \citet{Bautista2017}, and the cross correlation of Ly$\alpha$ forest and quasars from \citet{Font-Ribera2014}. These measurements provide both distance and Hubble parameter measurements at the effective redshift of their respective samples. 
Additional spherically-averaged distance measurements ($D_V$) are 6dFGS \citet{Beutler2011},
SDSS MGS \citet{Ross2015}, WiggleZ \citet{Kazin2014}, and eBOSS DR14 isotropic BAO \citet{DR14alphabetical}. 
In particular, the DR14 isotropic BAO result (labeld as `DR14-Iso' in Fig.~\ref{fig:planck_comparison}) analyzes the same sample as our work and reports a spherically-averaged distance measurement of $D_V(z=1.52) = 3843\pm147 \ ( r_{\rm d, fid} / r_{\rm d})$ Mpc.
As a comparison, we derive spherically averaged distance measurement from our $D_M$ and $H$ measurements at the same redshift and obtain $D_V(z=1.52) = 3871\pm157 \ (r_{\rm d} / r_{\rm d,fid})$ Mpc without redshift weighting and $3860 \pm 204 \ (r_{\rm d, fid} / r_{\rm d})$ Mpc with redshift weighting. These measurements are all consistent with the \citet{DR14alphabetical} measurement.    
In addition, we note that our Hubble parameter measurement spans a redshift range ($0.8 < z < 2.2$) that has not been measured in previous redshift surveys.

\section{Final Sample Outlook}
\label{sec:outlook}
The DR14 quasar sample covers 1192 deg$^2$ and 852 deg$^2$ of NGC and SGC regions. This solid angle is approximately a quarter of the final footprint of 7500 deg$^2$ for clustering quasars. The quadruple increase in footprint will result in reduced noise in the final sample. 
In this section, we assess the outlook of BAO measurements as would be obtained from the final eBOSS sample. 

To mimic the noise level in the final sample clustering quasars, we average the correlation functions from every four mock catalogues. This simple averaging serves to reflect the quadruple increase in footprint. After the averaging, we obtain 100 averaged mock correlation functions (labeled `4x' mocks) from the original 400 QPM mocks. We indeed observe greatly reduced noise in these `4x' mock correlation functions.

\begin{figure}
\includegraphics[width=0.48\textwidth]{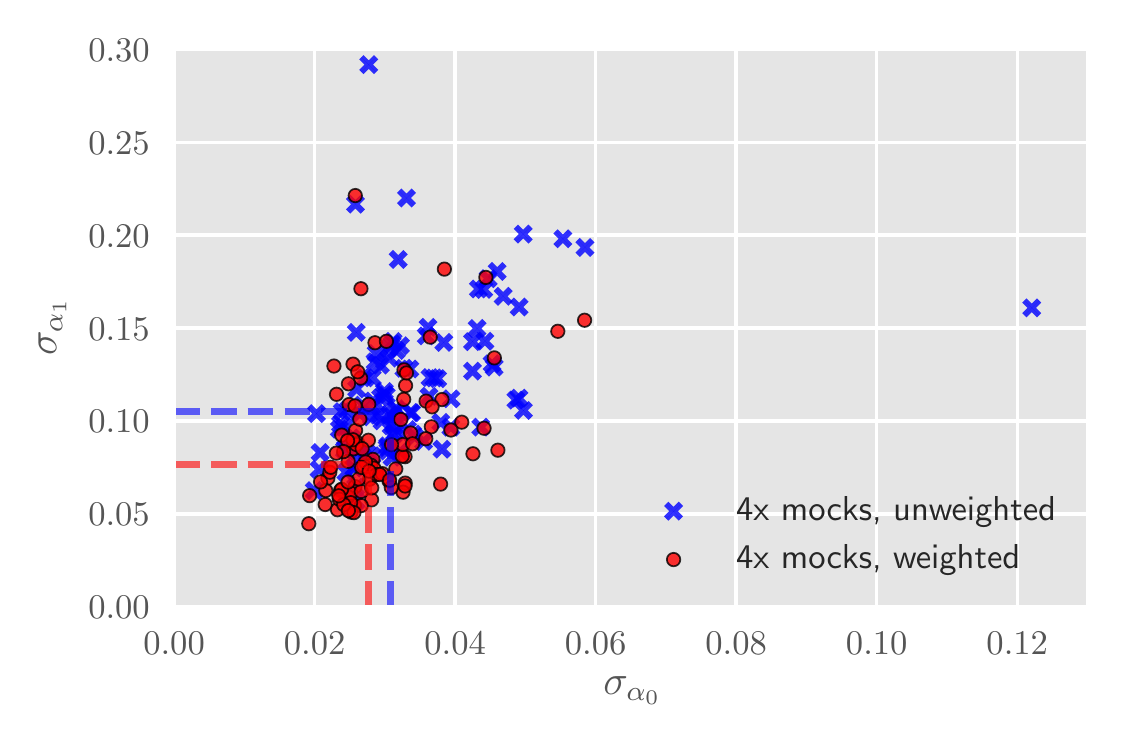}
\caption{The $\sigma_{\alpha_0}$ and $\sigma_{\alpha_1}$ values measured from the `4x' mocks. 
The measurements without redshift weighting are denoted by blue crosses, 
while the ones with redshift weighting are denoted by red circles. The vertical and horizontal dashed lines correspond to the error of inverse variance weighted mean of $\alpha_0$ and $\alpha_1$ from the mocks, multiplied by $\sqrt{100}$ for easy comparison with individual `4x' mock points.}
\label{fig:4x_mocks}
\end{figure}

We analyze the aforementioned 100 `4x' mock correlation functions with the same method 
outlined in the previous sections. 
The fitting results of these mocks are unbiased (see Table~\ref{tab:bestfit_tab}). 
Fig.~\ref{fig:4x_mocks} presents the errors $\sigma_{\alpha_0}$ and $\sigma_{\alpha_1}$ measured from the 100 `4x' mocks. 
We aggregate the mock measurements of $\alpha_0$ and $\alpha_1$ by calculating the inverse variance weighted average by
${\hat \alpha} = \frac{\sum \alpha_i/\sigma_{\alpha_i}^2}{\sum 1/\sigma_{\alpha_i}^2}$. The summation is over the 100 `4x' mocks. 
The error of ${\hat \alpha}$ is given by $\sigma({\hat \alpha}) = 1/\sqrt{\sum \frac{1}{\sigma_{\alpha_i}^2}}$.  We scale this error by $\sqrt{100}$ for ease of comparison with individual `4x' mock errors. 
The vertical and horizontal dashed lines in Fig.~\ref{fig:4x_mocks} show these statistics. 
The error $\sigma_{{\hat \alpha}_0}$ decreases from 3.1 per cent to 2.8 per cent. 
 Similarly, the weighted analysis gives an error of $\sigma_{{\hat \alpha}_1}$ of 7.7 per cent, compared to a 10.5 per cent without redshift weighting. 
 These results correspond to a 10 per cent improvement in $\alpha_0$ and a 27 per cent improvement 
in $\alpha_1$. 

\begin{figure}
\includegraphics[width=0.48\textwidth]{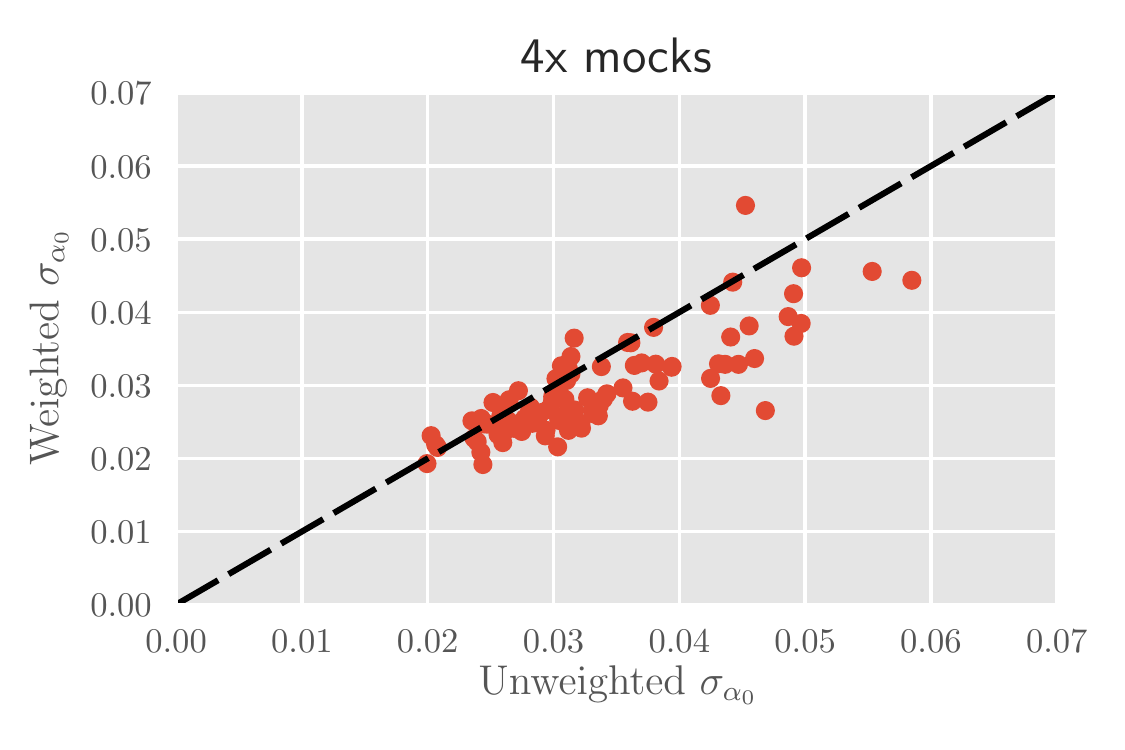}
\includegraphics[width=0.48\textwidth]{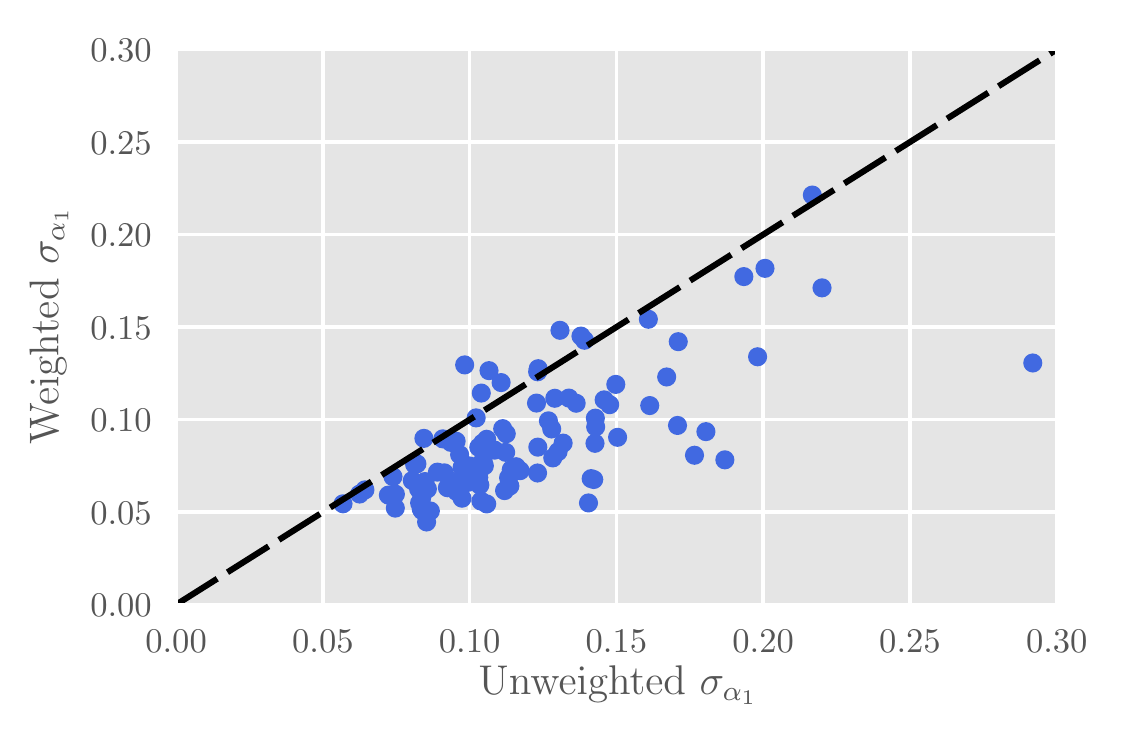}
\caption{The weighted and unweighted $\sigma_{\alpha_0}$ and $\sigma_{\alpha_1}$ values measured from the 100 `4x' mocks. The dashed line in the figure corresponds to a straight line of unit slope. The majority of points lie below the dashed line, suggesting redshift weighting is likely to be efficient for the final quasar sample. }
\label{fig:4x_mocks_comparison}
\end{figure}

Among the 100 `4x' mock measurements, 83 have an improved $\sigma_{\alpha_0}$ and 89 show an improved $\sigma_{\alpha_1}$ when we apply the redshift weights. This behavior can be clearly seen in 
Fig.~\ref{fig:4x_mocks_comparison}. 
The dashed line in the figure corresponds to a straight line of unit slope. 
The majority of points fall below this units line. 
Redshift weighting produces improved measurement errors for more than 80 per cent of the `4x' mocks, 
demonstrating that although redshift weighting does not yield smaller $\sigma_{\alpha_0}$ and $\sigma_{\alpha_1}$ for the current sample, it will likely be efficient for the final quasar sample.

The gains from redshift weighting in the `4x' mocks are much more significant than in the original QPM mocks. 
This result occurs because some mocks among the 400 individual QPM mocks are quite noisy 
and possess a weak BAO feature. 
As a result, these weak BAO detections lead to non-Gaussian likelihood surface. 
While redshift weighting is powerful at turning a `mediocre' measurement into a `good' one, it cannot turn a `bad' measurement (a non-detection of the BAO feature, for example) into a `mediocre' or `good' measurement. These noisy mocks thus render redshift weighting not as effective. After averaging, the `4x' mocks have better signal-to-noise ratio and enhanced BAO features. In fitting the `4x' mocks, the number of weak and non-detections is significantly reduced an  redshift weighting thus becomes much more efficient in tightening the error bars.
The substantial gains demonstrated in the `4x' mocks suggest redshift weighting will play an important role in unlocking the full potential of the BAO constraints from the final quasar sample.

\section{Discussion}

The DR14 quasar sample covers a wide redshift range from $z=0.8$ to 2.2. To analyze the BAO information in such a large range without sacrificing signal-to-noise ratio by splitting the sample into redshift slices, redshift weighting \citep{Zhu16} is a natural choice.
In this paper we have presented an anisotropic BAO analysis of the BOSS DR14 quasar sample using this technique

We approximate the distance-redshift relation, relative to
a fiducial model, by a quadratic function. By measuring the
coefficients from the mocks, we then reconstruct the distance
and Hubble parameter measurements from the expansion. Our
approach thus yields measurements of $D_M(z)$ and $H(z)$ at all
redshifts within the range of the sample. This approach differs from
previous analyses in which only measurements at the `effective
redshift' are given. 
We provide distance and Hubble parameter constraints at all redshifts within the redshift span of the sample. 

We first establish the effectiveness of redshift weighting in producing unbiased optimized constraints from a set of mock catalogues. With the same methodology, we analyze the BOSS DR14 quasar sample and achieve improved $D_M$ and $H$ constraints in fitting the BAO feature in the sample. Our $D_M$ error ranges from 4.6 per cent at $z = 2.2$ to 10.5 per cent at $z = 0.8$. Our $H$ error ranges from 4.6 per cent at $z = 0.8$ to 23.5 per cent at $z = 2.2$. 

To examine what will be possible when the final quasar sample becomes available, we generate a new set of mock catalogues with smaller noise by averaging every 4 of the original DR14 mocks to approximate the final eBOSS quasar sample. We analyze these averaged mocks with the same methodology and observe that redshift weighting offers significant improvement in the measurement errors over the single-bin analysis without redshift weighting. 
This demonstration suggests redshift weighting is important to unlocking the full BAO information within the sample. 

The power of redshift weighting lies in its optimal use of the information without splitting the sample into redshift slices. Although one can retain sensitivity to redshift by repeating traditional analyses on multiple slices and properly accounting for covariance between slices, this approach significantly adds to complexity of the analysis.

The method is especially useful when the survey covers a wide range of redshifts. 
Its success on the set of mock catalogues that mimic the final quasar sample shows promise that the method will be extremely useful for upcoming surveys like the Dark Energy Spectroscopic Instrument (DESI) \citep{DESI2016a, DESI2016b}. An anisotropic BAO analysis with similar redshift weighting techniques in Fourier space will appear in \citet{DDWang2018}. 
They optimise the measurements by deploying redshift weights constructed for the BAO signal 
in the quasar power spectrum. Different from how this work utilises the redshift weights, \citet{DDWang2018} assign the weights to individual quasars instead of weighting quasar pairs. 
Apart from this difference, the methodology is similar to \citet{Zhu16} and this work. 
Different from this work, \citet{DDWang2018} find applying redshift weighting on the DR14 sample
 produces improved measurement over the traditional single-bin analysis. 
This difference may be due to the difference in methodology and noise properties of the power spectrum and the correlation function. 
Despite this difference, the results reported in both works are fully consistent with each other within uncertainty. 
Besides these works, similar analysis methods inspired by the BAO redshift weights have been proposed to constrain redshift space distortions \citep{Ruggeri2016} and primordial non-Gaussianity \citep{Mueller2017} in upcoming surveys. RSD Measurements on the DR14 sample utilizing a similar methodology will appear in \citet{Ruggeri2018} and \citet{Zhao2018}. Redshift weighting can bring us closer to realizing the full capabilities of these surveys as we aim towards an ever increasing understanding of the expansion history of the universe.

\label{sec:discussion}

\section{Acknowledgments}

FZ would like to thank Tomomi Sunayama for helpful conversations. 
This work was supported in part by the National Science Foundation under 
Grant No. PHYS-1066293.
NP and FZ are supported in part by a DOE Early Career Grant DE-SC0008080. 
FB is a Royal Society University Research Fellow.

Funding for the Sloan Digital Sky Survey IV has been
provided by the Alfred P. Sloan Foundation, the U.S. Department
of Energy Office of Science, and the Participating
Institutions. SDSS-IV acknowledges support and resources
from the Center for High-Performance Computing at the
University of Utah. The SDSS web site is www.sdss.org.

SDSS-IV is managed by the Astrophysical Research
Consortium for the Participating Institutions of the SDSS
Collaboration including the Brazilian Participation Group,
the Carnegie Institution for Science, Carnegie Mellon University,
the Chilean Participation Group, the French Participation
Group, Harvard-Smithsonian Center for Astrophysics,
Instituto de Astrofísica de Canarias, The Johns
Hopkins University, Kavli Institute for the Physics and
Mathematics of the Universe (IPMU) / University of Tokyo,
Lawrence Berkeley National Laboratory, Leibniz Institut für
Astrophysik Potsdam (AIP), Max-Planck-Institut für Astronomie
(MPIA Heidelberg), Max-Planck-Institut für Astrophysik
(MPA Garching), Max-Planck-Institut für Extraterrestrische
Physik (MPE), National Astronomical Observatory
of China, New Mexico State University, New
York University, University of Notre Dame, Observatário
Nacional / MCTI, The Ohio State University, Pennsylvania
State University, Shanghai Astronomical Observatory,
United Kingdom Participation Group, Universidad Nacional
Autónoma de México, University of Arizona, University
of Colorado Boulder, University of Oxford, University of
Portsmouth, University of Utah, University of Virginia, University
of Washington, University of Wisconsin, Vanderbilt
University, and Yale University.

This research used resources of the National Energy Research Scientific Computing Center, 
a DOE Office of Science User Facility supported by the Office of Science of the U.S. Department of Energy under Contract No. DE-AC02-05CH11231.

\bibliographystyle{mnras}
\bibliography{qso_bao}

\end{document}